\documentclass[12pt]{article}
\usepackage{amsmath,epsfig,graphicx,amssymb}
\newcommand{\beq}{\begin{equation}}
\newcommand{\eeq}{\end{equation}}
\newcommand{\ben}{\begin{eqnarray}}
\newcommand{\een}{\end{eqnarray}}
\newcommand{\bes}{\begin{subequations}}
\newcommand{\ees}{\end{subequations}}
\newcommand{\bFig}{\begin{figure}}
\newcommand{\eFig}{\end{figure}}
\begin{document}

\title{A Relativistic Particle and Gravitoelectromagnetism}
\author{Partha Ghose\footnote{partha.ghose@gmail.com} \\
Centre for Natural Sciences and Philosophy, \\ 1/AF Bidhan Nagar,
Kolkata, 700 064, India\\
and\\Centre for Philosophy and Foundations of Science,\\Darshan Sadan, E-36 Panchshila Park,
 New Delhi 110017, India}
\maketitle

PACS Numbers 45.20.-d, 03.65.Ca, 04.20 Cv, 04.25 Nx
\vskip0.1in
\begin{abstract}
The mutual interaction of a relativistic particle and gravitoelectromagnetism (GEM) is studied both classically and quantum theoretically. 
\end{abstract}

\section{Introduction}

The purpose of this paper is to study the mutual interaction of a relativistic particle and gravitoelectromagnetism (GEM), usually presented as a post-Newtonian approximation of Einstein's theory of gravitation \cite{mashhoon1}, \cite{braginsky}, both classically and quantum theoretically. However, GEM can also be defined at the full nonlinear level of general relativity \cite{jantzenetal}. In order to proceed it is necessary first to study the classical theory of the system. We find new results in both the classical and the quantum theories. The latter are likely to have important implications for measurement theory which will be studied elsewhere. 

\section{Gravitoelectromagnetism (GEM)}
Let us start with a short review of gravitoelectromagnetism (GEM) in order to establish the notation and make the differences and analogies with ordinary electromagnetism clear. The local splitting of the general relativistic space-time manifold into `space' and `time' by means of observer congruence is the key to gravitoelectromagnetism in the general sense. It permits a better interface of our intuition based on our experience of 3-dimensional space and the pseudo-Riemannian geometry of 4-dimensional space-time. There are three space-time spittings that have been used in the literature, namely {\it slicing} or the 3 + 1 splitting popularised by Arnowitt, Deser and Misner \cite{adm}, {\it threading} or the 1 + 3 splitting introduced by Landau and Lifshitz \cite{landau} and {\it slicing plus threading} or the {\it nonlinear reference frame} \cite{jantzenetal}. A study of the relationships between these different splittings has led to a common mathematical framework and a `relativity of splitting formalisms' \cite{jantzenetal} which breaks down the artificial barriers dividing them and reveals complementary features of a common geometrical structure imposed on space-time ({\it the nonlinear frame}), offering a richer insight into them than any one of them on its own can offer. Generally, the lapse function and the shift vector in the metrics resulting from these splittings serve as the scalar and vector potentials for the gravitoelectric (GE) and gravitomagnetic (GM) force fields respectively. It is in this general sense that I will use the term `gravitoelectromagnetism' in this paper.

The analogy between GEM and Maxwell's equations is often presented in the literature for the weak and post-Newtonian gravitational field. In this approximation of the theory written using the 3 + 1 splitting of space-time and the family of harmonic gauges, the space-time metric around a rotating object is of the form

\ben ds^2 &=& g_{\mu\nu} dx^\mu dx^\nu \nonumber\\ &=&
- c^2 (1 - \frac{A_g^0}{c^2})\, d t^2 - \frac{4}{c}(A_{g i}dx_i) dt + (1 +
\frac{A_g^0}{c^2})\delta_{i j}dx_i dx_j.\label{metric}\een 
where $A_g^0$ and $A_{g i}$, the lapse function and the shift vector respectively, are identified with the scalar and vector potentials of GEM, and the GE and GM force fields are defined in terms of them by the equations

\ben {\bf E}_g
&=& - \nabla (\frac{1}{2}A_g^0) - \frac{1}{c}\frac{\partial}{\partial t} \left(\frac{1}{2}{\bf
A}_g \right),\\{\bf B}_g &=& \nabla \wedge {\bf A}_g.\een
To lowest order in these potentials $A_g$ the Lagrangian of a test particle in the GEM field can be shown to be \cite{mashhoon1} \ben L &=& - m c \frac{ds}{dt}\\\nonumber &=& -
m c^2 \sqrt{1 -
\beta^2} + \frac{1}{2} m \gamma (1 + \beta^2) A_g^0 - \frac{2m}{c}\gamma {\bf v} . {\bf A}_g + \cdots \label{lagrangian}\een where $\gamma$ is the Lorenz factor. It follows from this that the `kinetic' momentum of the test paticle is 
\beq
{\bf \pi} = {\bf p} - (2 m/c){\bf  A}_g \label{kinmom}\eeq 
where ${\bf p}$ is the canonical momentum \cite{note}. Since it is clear from (\ref{metric}) that the potentials $A_g$ characterise the space-time metric which is psudo-Riemannian, they are fundamentally different from the electromagnetic potentials $A$ which do not affect the flat space-time metric. This difference will be exploited in what follows to obtain new results. Eqns. (\ref{metric}) and (\ref{kinmom}) are the only results of importance from GEM that I will need for my purpose.

The interested reader will find the usual presentation of GEM in the weak field limit in Appendix A and recent developments in testing gravitomagnetism in the solar system in Mashhoon \cite{mashhoon2}. Unnikrishnan has claimed that large gravitomagnetism is 
inevitable in a matter-filled critical and flat Robertson-Walker universe and that it has
large effects on quantum systems and their phases, produced through
the modification ${\bf p} \rightarrow {\bf p} - (2 m/c) {\bf A}_g$
of their canonical momentum by the cosmological gravitomagnetic potentials
\cite{unni}.

It will be well to bear in mind before moving on to the next section that GEM {\it can} be defined more generally within full Einstein gravity.

\section{Classical Theory}

In all previous works on the effects of GEM on mechanical systems the
GEM potentials have been treated as external classical potentials.
This is an approximation. In this section we will study a relativistic system consisting of a
particle and GEM in mutual interaction, analogous to the interaction of a charged relativistic particle with the electromagnetic field, treating all the dynamical variables on the same footing.

Let us first consider the classical special-relativistic theory of a hypothetical free
particle. Its configuration space is Lorentzian and according to standard procedure (sketched in Appendix B), its momentum $p_\mu$ and coordinate $q_\mu$ satisfy the Poisson
brackets \beq \{q_\mu, p_\nu\}_{(q,p)} = \delta_{\mu \nu}.\label{pb1} \eeq
Similarly, in the absence of a test particle the GEM potential $A_{g \mu}$ has a
canonical conjugate momentum $\pi_{g \mu} = (1/c\, G) F_{g \mu \lambda}
\eta^\lambda d^3 q$ (see \cite{abraham} and the Appendix) satisfying the Poisson bracket \beq \{A_{g \mu}, \pi_{g \nu} \}_{(A_g,\pi_g)}  =
\delta_{\mu \nu}.\label{pbx}\eeq These two pairs form the set of four
independent canonical variables of the system we wish to study. The Hamiltonian of the system is0
\beq
H_0 = H_P (p) + H_{GEM} (A_g, \pi_g)
\eeq where the first term is the Hamiltonian of the free particle and the second term the GEM Hamiltonian. 

Since we are not actually interested in a hypothetical particle but in one that is in a GEM field, as we have seen in the previous section, in the weak field limit of general relativity its `kinetic' momentum $\pi_\mu$ is related to its canonical momentum $p_\mu$ by \beq \pi_\mu = p_\mu - \frac{2 m}{c}A_{g \mu}\label{canmom}\eeq with \beq {q_\mu, \pi_\nu}_{(q, \pi)} =
\delta_{\mu \nu}.\label{pb2}\eeq Eqn. (\ref{canmom}) is the momentum space representation (or pullback map) of the covariant derivative 
\beq
D_\mu = \partial_\mu - \frac{i}{2mG}A_{g \mu}\label{covderivative}
\eeq in the system's pseudo-Riemannian configuration manifold. This covariant derivative reveals an important geometrical feature of the GEM one-form $A_g = A_{g\mu}dx^\mu$, namely that it is also a connection one-form. As explained in Appendix B, there is a natural connection one-form that `connects' the tangent planes at two neighbouring points on the pseudo-Riemannian configuration manifold of the system along a curve. Eqns. (\ref{canmom}) and (\ref{covderivative}) tell us that $A_g$ is, in fact, that connection. Let the coordinates of the two neighbouring points connected by $A_g$ be $q$ and ${\cal{Q}}$ and let \beq {\cal{Q}}_\mu = q_\mu - \frac{c}{2m}\pi_{g \mu} \label{cancom2}\eeq with $\pi_{g \mu}$ connecting the two coordinates $q$ and ${\cal{Q}}$, just as $A_g$ connects the two momenta $p$ and $\pi$ (Eqn. \ref{canmom}) (see the Appendix for a differential geometric justification). It has the properties \beq
[{\cal{Q}}_\mu, {\cal{Q}}_\nu] = 0,
\label{pbQ}\eeq
\ben\{{\cal{Q}}_\mu,\pi_\nu\}_{((q,A_g),(p,\pi_g))} &=& \{q_\mu, p_\nu\}_{(q, p)} + \{\pi_{g \mu}, A_{g \nu}\}_{(A_g, \pi_g)} = 0,\nonumber\\\{{\cal{Q}}_\mu,\pi_\nu\}_{((q,A_g),(\pi,\pi_g))}&=& 0\label{A1}\een and \ben \{{\cal{Q}}_\mu, p_\nu \}_{((q,A_g), (p, \pi_g))} &=& \{ q_\mu, p_\nu\}_{(q,p)} - \{\frac{c}{2 m}\pi_{g \mu}, p_\nu  \}_{(A_g,\pi_g)}= \delta_{\mu \nu},\nonumber\\
\{{\cal{Q}}_\mu, p_\nu \}_{((q,A_g),(\pi,\pi_g))} &=& \delta_{\mu \nu}.\label{pb3} \een It follows that in addition to $(q, \pi)$ one can also choose $({\cal{Q}}, p)$ as a canonical pair, and have
\beq
\{{\cal{Q}}_\mu,\pi_\nu\}_{(({\cal{Q}},A_g),(p,\pi_g))}= 0 \label{pb4}.
\eeq
Just as $\pi$ is the kinematic momentum of the particle conjugate to its canonical position $q$, the Poisson bracket (\ref{pb3}) implies that ${\cal{Q}}$
is the kinematic position conjugate to its canonical momentum $p$. ${\cal{Q}}$ and $\pi$ carry global information about the configuration manifold of the particle through the connection $A_g$ and its conjugate $\pi_g$. The relation (\ref{cancom2}) is usually ignored in considering mechanical systems in external  potentials. {\it The important new feature is that the Poisson brackets (\ref{A1}) vanish}. In the absence of GEM, which is locally possible (the equivalence principle), ${\cal{Q}} = q$ and $\pi = p$.

The total Hamiltonian of the interacting system is therefore
\beq H = H_P (p - (2m/c)A_g = \pi) + H_{GEM} (A_g, \pi_g - (2m/c)q = - (2m/c){\cal{Q}})\eeq.
The equations of motion for the particle in terms of the variables $({\cal{Q}}, p)$ are then \beq
\dot{{\cal{Q}}}_\mu = \{{\cal{Q}}_\mu , H \}_{({\cal{Q}},p)} = \frac{\partial
H}{\partial p^\mu}, \,\,\,\,\,\,\dot{p}_\mu
= \{p_ \mu , H\}_{({\cal{Q}},p)} = - \frac{\partial H}{\partial
{\cal{Q}}^\mu},\label{HE1}\eeq
and in terms of the variables $(q, \pi)$, they are 
\beq \dot{q_\mu} = \{ q_\mu , H \}_{(q,\pi)} =
\frac{\partial H}{\partial \pi^\mu}, \,\,\,\,\,\,\dot{\pi}_\mu =
\{\pi_\mu , H\}_{(q,\pi)} = - \frac{\partial H}{\partial
q^\mu}.\label{HE2}\eeq These equations show that $(q, p)$ act as
the fundamental canonical variables ensuring Hamiltonian flows underlying the
evolution of the kinematic variables $({\cal{Q}},\pi)$. The ${\cal{Q}}$-$\pi$ space
of the particle is a projection of a higher dimensional phase space $V_T$---it is dual to the canonical phase space $p$-$q$ (see Appendix).

The equations of motion for the GEM variables are
\beq
\dot{A}_{g \mu} = \{ A_{g \mu} , H \}_{(A_g,\pi_g)} =
\frac{\delta H}{\delta \pi_g^\mu}, \,\,\,\,\,\,\dot{\pi}_{g \mu} =
\{\pi_{g \mu} , H\}_{(A_g,\pi_g)} = - \frac{\delta H}{\delta
A_g^\mu}.
\eeq

Before ending this section let us note an important and well-known property of covariant derivatives, namely that they do not commute,
\beq
[D_\mu, D_\nu] = - \frac{i}{2mG}F_{g \mu\nu},\label{hol}
\eeq the commutator being the curvature of the connection. This non-commutativity of the covariant derivatives is a {\it classical} (i.e., non-quantum theoretic) result following from the geometrical fact that parallel transporting a vector around a closed loop on a curved manifold results in a different vector. This failure to return to the initial vector is known as holonomy. Thus, the covariant derivatives carry global information about the manifold.

\section{Quantum Theory}

It is now straightforward to construct the quantum theory of the system
by adopting the standard
canonical procedure of replacing the classical Poisson brackets
(\ref{pb1}) through (\ref{pb4}) by commutators. One gets \beq
[q_\mu, \hat{\pi}_\nu] = [q_\mu, \hat{p}_\nu] = i\hbar\, \delta_{\mu \nu},\label{comm1}\eeq \beq [A_{g \mu}, \hat{\pi}_{g \nu}] = i \hbar\, \delta_{\mu\nu},\label{comm2}\eeq
\beq[\hat{{\cal{Q}}}_\mu,\hat{\pi}_\nu] = [q_\mu,\hat{p}_\nu] + [\hat{\pi}_{g \mu} , A_{g \nu}] = 0,\label{A2}\eeq \beq [\hat
{{\cal{Q}}}_ \mu, \hat{p}_\nu] = i \hbar\, \delta_{\mu\nu},\label{comm3}\eeq
It follows from these commutators that $\hat{p}_\mu = - i\hbar\partial_\mu$, $\hat{\pi}_{g \mu} = - i \hbar \delta/\delta A_g^\mu$, $q_\mu$, $\hat{{\cal{Q}}}_ \mu$ and $\hat{\pi}_\mu$ are all hermitian. Hence, $\hat{\pi}_\mu = - i\hbar\partial_\mu - (2m/c) A_{g \mu}$ and therefore
\beq
[\hat{\pi}_\mu, \hat{\pi}_\nu] = \frac{2 i m \hbar}{c} F_{g \mu \nu} \label{picomm}
\eeq but
\beq
[\hat{{\cal{Q}}}_ \mu, \hat{{\cal{Q}}}_ \nu] = 0.\label{Qcomm}
\eeq
A comparison of (\ref{hol}) and (\ref{picomm}) shows that the latter is a consequence of the curvature and holonomy of the connection $A_g$. Thus, this commutator carries global information about the configuration manifold. For example, it vanishes in flat space-time regions where $F_{g\mu\nu} = 0$ but $A_{g\mu} \neq 0$ and $\hbar \neq 0$.

As in the classical theory, $\hat{{\cal{Q}}}$ and $\hat{\pi}$ are the kinematic coordinate and momentum operators respectively of the interacting particle which carry global information about the manifold, whereas $\hat{p}$ and $\hat{q}$ are the local canonical momentum and coordinate operators respectively that enable underlying Hamiltonian (or Schr\"{o}dinger) evolutions to occur. {\em Significantly, $\hat{{\cal{Q}}}$ and $\hat{\pi}$ have simultaneous eigenvalues because of (\ref{A2})}. This implies that quantum theory admits `trajectories' of the particle in the ${\cal{Q}}$-$\pi$ space. One can define the density operator

\beq
\hat{\rho} = \sum_j p_j \vert \psi_j \rangle \langle \psi_j\vert, \,\,\,\,\,\,\,\,\,\,\,\,\ p_j \geq 0, \,\,\,\,\,\,\,\,\,\,\,\,\sum_j p_j = 1
\eeq
with $\vert \psi_j \rangle$ forming a complete set of states that are simultaneous eigenstates of $\hat{{\cal{Q}}}$ and $\hat{\pi}$. It satisfies the evolution equation
\beq
i \hbar \frac{\partial \hat{\rho}}{\partial t} = [H, \hat{\rho}].
\eeq
Hence, one can define the density function of the trajectories as
\beq
G({\cal{Q}}, \pi, t) = \langle {\cal{Q}}, \pi\vert \hat{\rho}(t)\vert {\cal{Q}}, \pi\rangle
\eeq with \beq
\hat{\rho}(t) = U (t) \hat{\rho}(0) U^\dagger (t),\,\,\,\,\,\,\,\,\,\, U (t) = e^{- i H t/\hbar}.
\eeq

These new results raise important questions regarding the measurement problem and the physical significance of ${\cal{Q}}$ and $\pi$, which will be studied further and reported elsewhere. Here we would only like to point out that despite the existence of `trajectories' in the theory, there is a fundamental difference from the Bohm theory \cite{bohm}. The Bohm theory imposes an additional {\it ad hoc} condition, the guidance condition, on standard quantum mechanics to define trajectories in configuration space. When a particle is coupled to GEM and the total system is quantized, trajectories in the ${\cal{Q}}$-$ \pi$ space are inevitable consequences---no guidance conditions are required and there are no configuration space particle trajectories as
`hidden variables'. Furthermore, the `guidance condition' in the Bohm heory results in trajectories in cofiguration space whose initial distribution must be {\it chosen} to be identical with the quantum mechanical distribution. The continuity equation then ensures that this identity is preserved in time. The trajectories in $\cal{Q}$-$\pi$ space
are consequences of the commutation relations (\ref{A2}) which
are preserved in time, and the distribution of the trajectories is automatically determined by the theory for all times.

The formalism reduces to that of standard quantum mechanics of a free particle
in the {\it formal} limit of the GEM potentials vanishing.

\section{Concluding Remarks}

We have studied the case of a relativistic particle in a weak gravitoelectromagnetic field, treating the GEM and particle dynamical variables on a symmetrical footing. The kinematic momentum ${\bf \pi} = {\bf p} - (2 m/c){\bf  A}_g$ of the particle implies the introduction of a new canonical coordinate $Q$ and new Poisson brackets (\ref{pb1} - \ref{pb3}) and commutators (\ref{comm1} - \ref{Qcomm}) whose implications have not been explored before.

Although the GEM potential $A_g$ looks similar in many respects to other gauge potentials in physics, there is a fundamental difference between them, namely $A_g$ is the connection on the tangent bundle of a pseudo-Riemannian manifold whereas the other gauge potentials are connections on principal bundles on flat Lorentzian manifolds. As explained in Appendix B, this is the basis of introducing a new canonical variable $Q$ in classical theory.

The important question that remains unanswered is: what happens when
the non-linearities of General Relativity are fully taken into account?
GEM is often presented as the weak and linearised limit of GR involving a harmonic $U(1)$ gauge group whereas full GR looks like a Yang-Mills theory in terms of the Ashtekar variables. The question arises as to what would happen if the considerations described above in the paper could be generalized to full GR. The fact that GEM can also be defined at the full nonlinear level of general relativity \cite{jantzenetal} raises some hope, but technical difficulties abound.

\section{Acknowledgements}
I would like to thank C. S. Unnikrishnan for an invitation to visit
the Gravitation Group in the Tata Institute of Fundamental Research,
Mumbai and introducing me to GEM as well as his Cosmic Relativity
theory. The present paper has its origin in those stimulating
discussions, but he cannot be held responsible for any deficiencies
in it. I am also grateful to Dilip Bhattacharya for many helpful
discussions on manifolds and differential geometry. I am also indebted to him, Virendra Singh, C. S. Unnikrishnan and Archan Majumdar for many helpful comments on earlier drafts of the
paper which have been substantially revised as a
result.

\section{Appendix A}

Using the metric (\ref{metric}) Einstein's equations can be shown to reduce to the form
\cite{mashhoon1}

\ben
\partial_\mu F_g^{\mu\nu} &=&  4\pi G j^\nu,\label{maxwell1}\\
2 F_g^{\mu\nu} &=& \partial^\mu A^\nu_g - \partial^\nu A^\mu_g.\label{maxwell2}\een
Because these equations are `formally' analogous to Maxwell's quations of electrodynamics (but not isomorphic to them),  $F_g^{\mu\nu}$ is called the GEM Faraday tensor, $A^\mu_g =
(A^0_g, {\bf A}_g)$ the gravito-electromagnetic potential, $j^\nu =
(\rho, {\bf j}\,_g)$, $\rho=T^{00}/c^2$ the ``mass-charge'' density
so that $\int \rho d^3 x = M$, $M$ being the total mass, ${\bf j}$
the mass-current density and $A^0_g = 2\Phi$, $\Phi$ being
the Newtonian potential. One also has the Lorenz condition \beq
\partial_\mu A^\mu_g = 0.\label{lorenz}
\eeq The equations (\ref{maxwell2}) define the gravito-electric and
gravito-magnetic fields ${\bf E}_g$ and ${\bf B}_g$: \ben {\bf E}_g
&=& - \nabla \Phi - \frac{1}{c}\frac{\partial}{\partial t} \left(\frac{1}{2}{\bf
A}_g \right)\\{\bf B}_g &=& \nabla \wedge {\bf A}_g.\een The
equations (\ref{maxwell1}) and (\ref{lorenz}) can be written in
terms of these fields and the Newtonian potential $\Phi$ as
\ben \nabla\, .\, {\bf E}_g &=& 4\pi G \rho\\
\nabla\, .\, \frac{1}{2}{\bf B}_g &=& 0\\ \nabla \wedge  {\bf E}_g &=& -
\frac{1}{c}\frac{\partial}{\partial t}\left(\frac{1}{2} {\bf
B}_g\right)\\
\nabla \wedge (\frac{1}{2} {\bf B}_g) &=& \frac{1}{c}\frac{\partial
{\bf E}_g }{\partial t} + \frac{4\pi G}{c} {\bf j}\\
\frac{1}{c} \frac{\partial \Phi}{\partial t} + \nabla \,.\,
(\frac{1}{2}{\bf A}_g) &=& 0\een These equations describe the weak
gravitational field around a rotating object. 

Equations (\ref{maxwell1}), (\ref{maxwell2}) and (\ref{lorenz})
are invariant under the local gauge transformations \beq A^\mu_g
(x) \rightarrow A^\mu_g (x)+
\partial^\mu \chi (x),\,\,\,\,\,\,\Box \chi (x) = 0. \eeq
However, one can also discuss the effect of spatial gauge transformations on the GE and GM vector fields in a parametrized {\it nonlinear reference frame} instead of
in 3 + 1 splitting or {\it slicing}. In the {\it threading} point of view
the GE and GM fields are invariant, leading
to a gauge freedom analogous to that of the scalar and vector potentials for
the electric and magnetic fields (\cite{jantzenetal}, section 10).
 
\section{Appendix B}
In this section we will briefly summarise certain useful results in differential geometry to establish the notation and then justify the new results in the text in a more geometric fashion. Since GEM is involved, it becomes necessary to formulate the canonical approach in a form that manifestly preserves all relevant symmetries. This can be done using manifolds and differential geometry. A classical phase space is then defined as the space of solutions of the classical equations. One can always, if one wishes, choose a coordinate system with a time coordinate and identify the classical solutions with the initial data in that coordinate system, but there is no necessity to make such a non-covariant choice. The notion of a `symplectic structure on phase space' is a more intrinsic concept than the idea of choosing coordinates $q_i$ and $p_i$ \cite{witten}.

Consider the configuration space of a classical system which is
generally a manifold ${\cal{M}}$ with local charts $(U, x),\, x(m, m \in U)
= q = (q_1, q_2, \cdots, q_n) \in {\bf R}^n$. One can define the tangent
vectors $X_i^q = \partial/\partial q^i, \,q \in {\bf R}^n$ and via the inverse mapping $x^{- 1}$ the tangent vectors
$X_i^m = \partial/\partial x^i, m \in U$. These tangent vectors
span the tangent space at the point $m \in U$ and are fibres
on ${\cal{M}}$. The fibres on all points on ${\cal{M}}$ together
with ${\cal{M}}$ constitute the tangent bundle $T{\cal{M}}$. The
dual to the tangent bundle is called the cotangent bundle
$T^*{\cal{M}}$ with $\pi: T^*{\cal{M}}\rightarrow {\cal{M}}$ the projection.
One can define a canonical one-form $\theta$ on $T^*{\cal{M}}$ by
\beq
\theta (\alpha)w = \alpha. T \pi (w)
\eeq
where $\alpha \in T^*{\cal{M}}$ and $w \in T_\alpha (T^*{\cal{M}})$.
The canonical two-form is defined by $\omega = - d \theta,\, d \omega = 0$. This is a reflection of the fact that $T^*{\cal{M}}$ is a symplectic manifold. If ${\cal{M}}$ is finite dimensional, the formula for $\theta$ in a local chart $(U, x)$ may be written as
$\theta = \sum_i p_i dq^i$ where the exterior derivatives $dq^i$ span the
cotangent space and are dual to $X_i^q$: $\langle dq^i, X_j^q\rangle =
\delta^i_ j$. The $p_i$ are the momenta conjugate to the coordinates
$q_i$. The two-form $\omega (q,p) = - d
\theta = \sum_i d q^i \wedge d p_i$ and it is
closed, i.e., $d \omega = 0$. It is well-known that one can always
associate a Poisson manifold $(T^*{\cal{M}},\{,\})$ with the
sympletic manifold $T^*{\cal{M}}$. The fundamental Poisson brackets
of $q_i$ and $p_j$ in a chart $(U, x)$ are \beq \{q_i,p_j\} = \delta_{i j}.\eeq 
$T^*{\cal{M}}$ can be regarded as a $2n$ dimensional manifold called
`phase space' with coordinates $(q_1, \cdots, q_n, p_1,
\cdots, p_n) \in U$ rather than a bundle.

Similarly, for infinite dimensional systems like fields one
considers the manifold ${\cal{A}}_g$ of potentials $A_{g \mu}$. The
corresponding phase space is then the cotangent bundle
$T^*{\cal{A}}_g$ with the canonical symplectic structure. Since the
Lagrangian is \beq {\cal{L}}= - \frac{1}{4 G}F_{g \mu \nu}F_g^{\mu
\nu}, \eeq the canonical momentum is $\pi_{g \mu} 
=(1/c\, G) F_{g \mu 0} d^3 x = (1/c\, G) F_{g \mu \lambda}\eta^\lambda d^3 x$
with $\eta^\lambda\eta_\lambda = -1$. The canonical symplectic structure $\omega_g$ on
$T^*{\cal{A}}_g$ is
\beq
\omega_g \left((A_{g 1},\pi_{g 1}), (A_{g 2},\pi_{g 2})\right) = \int_{{\bf R}^3} (\pi_{g 2}.A_{g 1} - \pi_{g 1}.A_{g 2}) d^3 x,\eeq and the associated fundamental Poisson
bracket is \beq \{F,G \}_{(A_g, \pi_g)}
= \int_{{\bf R}^3} \left(\frac{\delta F}{\delta A_g}.\frac{\delta G}{\delta \pi_g} - \frac{\delta F}{\delta \pi_g}.\frac{\delta G }{\delta A_g} \right) d^3 x \eeq where $\delta F/\delta A_g$ is the vector field defined by
\beq
D_{A_{g}} F(A_g, \pi_g)\, .\, A^\prime = \int \frac{\delta F}{\delta A_g}\, . \, A^\prime  d^3 x\eeq with the vector field $\delta F/\delta \pi_g$ defined similarly. For further details, see \cite{abraham}.

\begin{figure}
\begin{picture}(100,100)
\put(95,45){$T_m M$}
\put(120,50){\vector(1,0){100}}
\put(165,40){$d \varphi_m$}
\put(225,45){$T_{m^\prime} M$}
\put(115,115){\vector(0,-2){60}}
\put(230,85){$\tau_{m^\prime}$}
\put(225,120){$T^*_{m^\prime} M$}
\put(120,120){\vector(1,0){100}}
\put(165,125){$d \varphi_m^*$}
\put(95,120){$T^*_m M$}
\put(225,115){\vector(0,-2){60}}
\put(100,85){$\tau_m$}
\end{picture}
\caption{The commutative diagram illustrating the pull-back map $d\varphi^*_m$}
\end{figure}

If one considers a test particle in GEM, the system manifold is not Lorentzian but a `curved' pseudo-Riemannian manifold ${\cal{M}}_{(A_g)}$. The metric (\ref{metric}) clearly shows the nontrivial and non-Lorentzian character of the space-time manifold on which the particle moves. The fact that a manifold $M$ is curved means that the tangent spaces $T_m M$ and $T_{m^\prime}M$ at two infinitesimally separated neighbouring points $m, m^\prime \in M$ are disjoint. The connection is a mapping of these tangent spaces. Let $\varphi: m \rightarrow m^\prime$ be a map. Then $d \varphi_m : T_m M \rightarrow T_{\varphi (m)} M$, i.e., the tangent vectors to $M$ at $m$ are mapped to the tangent vectors to $M$ at $m^\prime$ by the differential or covariant derivative $d \varphi_m$ which consists of the ordinary partial derivative plus the connection. This is the connection map $d \varphi_m$ (Fig. 1). Let $\tau_m: T^*_m M \rightarrow T_m M$ and
$\tau_{m^\prime}: T^*_{m^\prime} M \rightarrow T_{m^\prime} M$. Then $d \varphi_m^*: T^*_ m M \rightarrow T^*_{m^\prime} M$ is the pullback map. In terms of local charts on ${\cal{M}}_{(A_g)}$ let the map $p \rightarrow \pi = p - (2 m/c) A_g$ correspond to the pullback map $d \varphi_m^*$, and let the coordinates of the $T^*_m M$ bundle be $(\cal{Q}, \pi)$. Since $(q,p)$ and $(A_g, \pi_g)$ are canonical pairs, we have ${\cal{Q}} = q - (c/2 m) \pi_g$. This is the differential geometric justification for Eqn. (\ref{cancom2}) in the text.

One can regard $(q, {\cal{Q}})
\in V \times V$ where $V$ is a vector space and $(p\,
\otimes \pi) \in V \times V$. Then one has a nondegenerate
symplectic two-form on $V \times V \times V \times V$ given by
\ben \omega \left((q, {\cal{Q}})_1,
(p \otimes \pi)_1, (q, {\cal{Q}})_2, (p \otimes
\pi)_2\right) &=& (p_1 \otimes \pi_1)
(q_1, {\cal{Q}}_1) - (p_2 \otimes \pi_2) (q_2, {\cal{Q}}_2)\nonumber\\
&=& \pi_1 (q_1)\,.\, p_ 1 ({\cal{Q}}_1) - \pi_2 (q_2)\,.\, p_
2 ({\cal{Q}}_2). \een
Consider the phase spaces $({\cal{Q}}, p) \in V_{{\cal{Q}}} \times V_{{\cal{Q}}}^*$ and $(q, \pi) \in V_q \times V_q^*$. The total space is $V_T = V_{{\cal{Q}}} \times V_{{\cal{Q}}}^* \times V_q \times V_q^*$. Now consider the projections $P_1 V_T = V_{{\cal{Q}}} \times V_{{\cal{Q}}}^* \times V_q^* \equiv V^\prime$ and $P_2 V^\prime = V_{{\cal{Q}}} \times V_q^*$. Then $P_2 P_1 V_T  = V_{{\cal{Q}}} \times V_q^*$, and $({\cal{Q}}, \pi) \in V_{{\cal{Q}}} \times V_q^*$. This shows that the space ${\cal{Q}}$-$\pi$ is a projection of the higher dimensional phase space $V_T$ that allows Hamiltonian flows. Similarly, consider the projections $P_3 V_T = V^*_{{\cal{Q}}} \times V_{q} \times V_q^* \equiv V^{\prime \prime}$ and $P_4 V^{\prime \prime} = V_{{\cal{Q}}}^* \times V_q$. Then $P_4 P_3 V_T  = V_{{\cal{Q}}}^* \times V_q$, and $(p, q) \in V_{{\cal{Q}}}^* \times V_q$, which shows that the space $p$-$q$ is another projection of $V_T$ whose dual is the space ${\cal{Q}}$-$\pi$.

A word is in order about the essential difference between the GEM potential $A_g$ and other gauge connections like the electromagnetic and Yang-Mills potentials. As already mentioned, $A_g$ connects the tangent spaces at two neighbouring points $m$ and $m^\prime$ on the pseudo-Riemannian manifold ${\cal{M}}_{(A_g)}$. Other gauge potentials are connections on `principal bundles' $P$ with the elements of some internal Lie group $G$ as fibres on a flat space-time base manifold $\cal{M}$. The curvature represented by the tensor $F = F_{\mu \nu}^a T^a d x^\mu \wedge d x^\nu$ in such gauge theories ($T^a$ being the generators of $G$) refers to the curvature of the principal bundle manifold. The disjoint tangent spaces involved in these theories are tangent spaces to $P$ at neighbouring points on it, but the tangent spaces to the flat base manifold $\cal{M}$ are essentially the same everywhere and they can always be mapped by a simple `change of coordinates'. Hence, no analogue of Eqn. (\ref{cancom2}) is called for when considering these gauge connections.


\begin{thebibliography}{0}
\bibitem{note}
These definitions of `canonical' and `kinetic' momenta are different from those used in Ref. 1.
\end{thebibliography}
\end{document}